\documentclass[
prl,onecolumn,
amsfonts,amsmath,amssymb, 
showpacs,
superscriptaddress]{revtex4}

\usepackage{graphicx}
\usepackage{amsmath}
\usepackage[mathscr]{eucal}

\newcommand{\Y}{\Delta\mathcal{S}}

\begin{document}

\title{Dynamic Relaxation of a Levitated Nanoparticle from a Non-Equilibrium Steady State}

\author{
 Jan Gieseler
 }
 \affiliation{ICFO-Institut de Ciencies Fotoniques, Mediterranean Technology Park,
 08860 Castelldefels (Barcelona), Spain}

\author{
Romain Quidant
}
 \affiliation{ICFO-Institut de Ciencies Fotoniques, Mediterranean Technology Park,
 08860 Castelldefels (Barcelona), Spain}
 \affiliation{ICREA-Instituci{\'o} Catalana de Recerca i Estudis Avan\c{c}ats, 08010 Barcelona, Spain}

\author{
Christoph Dellago
}
 \affiliation{University of Vienna, Faculty of Physics, Boltzmanngasse 5, 1090 Wien, Austria}

\author{
Lukas Novotny
}
 \affiliation{Photonics Laboratory, ETH Z\"urich, 8093 Z\"urich, Switzerland}


\vspace{-3.2em}

\begin{abstract}
Fluctuation theorems are a generalization of thermodynamics on small scales and provide the tools to characterise the fluctuations of thermodynamic quantities in non-equilibrium nanoscale systems. They are particularly important for understanding irreversibility and the second law in fundamental chemical and biological processes that are actively driven, thus operating far from thermal equilibrium. 
Here, we apply the framework of fluctuation theorems to investigate the important case of a system relaxing from a non-equilibrium state towards equilibrium. Using a vacuum-trapped nanoparticle, we demonstrate experimentally the validity of a fluctuation theorem for the relative entropy change occurring during relaxation from a non-equilibrium steady state. 
The platform established here allows non-equilibrium fluctuation theorems to be studied experimentally for arbitrary steady states and can be extended to investigate quantum fluctuation theorems as well as systems that do not obey detailed balance. 
\end{abstract}

\maketitle


\noindent One of the tenets of statistical physics is the central limit theorem. It allows systems with many microscopic degrees of freedom to be reduced to only a few macroscopic thermodynamic variables. The central limit theorem states that, independently of the distribution of the microscopic variables, a macroscopic extensive quantity $U$, such as the total energy of a system with $N$ degrees of freedom, follows a Gaussian distribution with mean $\langle U\rangle\propto N$ and variance $\sigma^2_U \propto N$. Consequently, for large $N$, the relative fluctuations $\sigma_U/\langle U\rangle$ vanish and the macroscopic quantity becomes sharp. With the advance of nanotechnology it is now possible to study experimentally systems small enough such that the relative fluctuations become comparable to the mean value. This gives rise to new physics where transient fluctuations may run counter to the expectations of the second law of thermodynamics~\cite{Wang:2002hw}.\\[-2ex]

The statistical properties of the fluctuations of thermodynamic quantities like heat, work and entropy production are described by exact relations known as fluctuation theorems \cite{Crooks:1999wq,Jarzynski:1997uj,Bochkov:1981ua,Machlup:1953wa}, which permit to express the inequalities familiar from macroscopic thermodynamics as equalities \cite{Jarzynski:2011hl,Seifert2012}.
Fluctuation relations are particularly important for understanding fundamental chemical and biological processes, which occur on the mesoscale where the dynamics are dominated by thermal fluctuations \cite{Bustamante:2005tg}. They allow us, for instance, to relate the work along non-equilibrium trajectories to thermodynamic free-energy differences \cite{Alemany:2012jo,Collin:2005fx}. Fluctuation theorems have been experimentally tested on a variety of systems including pendulums \cite{Douarche:2006im}, trapped microspheres \cite{Wang:2002hw}, electric circuits \cite{Garnier:2005iu}, electron tunneling \cite{Kung:2012ct,Saira:2012bd}, two-level systems \cite{Schuler:2005eu} and single molecules \cite{Hummer:2001ts,Liphardt:2002ui}. Most of these experiments are described by an overdamped Langevin equation. However, systems in the underdamped regime \cite{Ciliberto:2010jg}, or in quantum systems \cite{Campisi:2011ka} where the concept of a classical trajectory looses its  meaning, are less explored.\\[-2ex]

Here, we study  the thermal relaxation of a highly underdamped nanomechanical oscillator from a non-equilibrium steady state towards equilibrium. Because of the low damping of our system, the dynamics can be precisely controlled even at the quantum level~\cite{Teufel:2012jg,Chan:2011dy,OConnell:2010br}. This high level of control allows us to produce non-thermal steady states and makes nanomechanical oscillators ideal candidates for investigating non-equilibrium fluctuations for transitions between arbitrary steady states. While for the initial steady state detailed balance is violated, the relaxation dynamics are described by a microscopically reversible Langevin equation that satisfies detailed balance \cite{Crooks:JStatMech:2011dx}. Under these conditions, a transient fluctuation relation holds \cite{Evans:2002gy,Seifert2012} for the relative entropy change characterising the irreversibility of the relaxation process. Similar relations hold also for relaxation processes in ageing systems as studied both theoretically \cite{Crisanti:2007gi} and experimentally \cite{GomezSolano:2012gi,GomezSolano:2011em,Ciliberto:2013fc} in gels and glasses. For the initial non-equilibrium steady state generated in our experiment we derive an analytical expression for the phase space distribution, which is in excellent agreement with the experimental data and directly validates the fluctuation theorem. Our experimental framework can be extended to  study transitions between arbitrary steady states and, furthermore, lends itself to the experimental investigation of quantum fluctuation theorems~\cite{Huber:2008go} for nanomechanical oscillators~\cite{Teufel:2012jg,Chan:2011dy,OConnell:2010br}.\\[-2ex]

\begin{figure}[!t]
\includegraphics[width=0.75\textwidth]{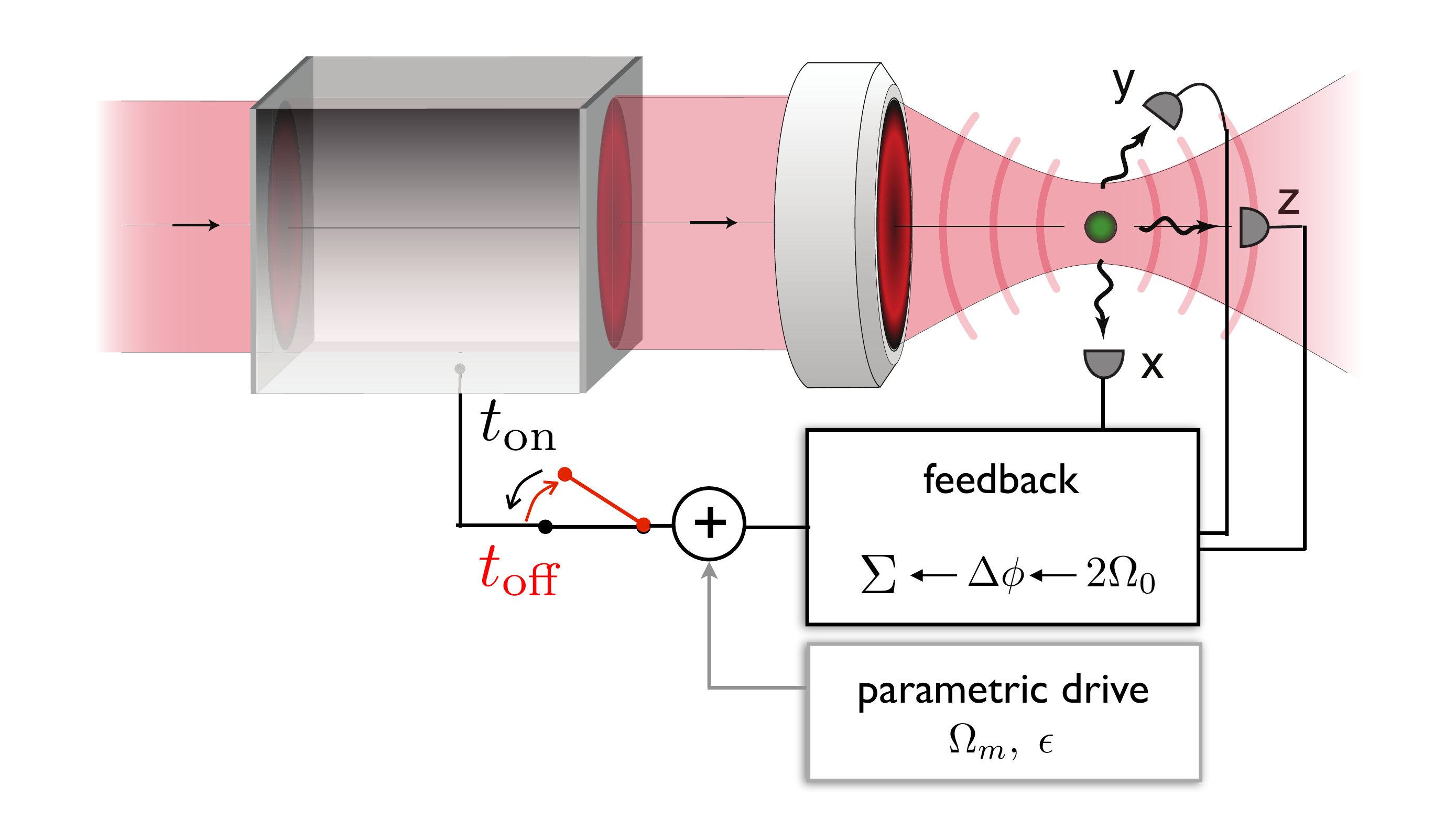}
\vspace{-1em}
\caption{
Experimental setup.
A nanoparticle is trapped by a tightly focused laser beam in high vacuum.
In a first experiment, the nanoparticle is initially cooled by parametric feedback.
At time $t=t_{\rm off}$ the feedback is switched off and the nanoparticle trajectory is followed as it relaxes to equilibrium.
After relaxation, the feedback is switched on again and the experiment is repeated.
In a second experiment the nanoparticle is initially excited by an external modulation in addition to feedback cooling.
Again at a time $t=t_{\rm off}$, both the feedback and the external modulation are switched off and the nanoparticle is monitored as it relaxes.
\label{fig:Figure0}}
\end{figure}
%
The experimental setup is shown in Fig.~\ref{fig:Figure0}. We consider a silica nanoparticle of radius $r\sim 75\,$nm and mass $m\sim 3\times 10^ {-18}$\,kg that is trapped in vacuum by the gradient force of a focused laser beam. Within the trap, the nanoparticle oscillates in all three spatial directions. To first approximation, the three motional degrees of freedom are well decoupled. Hence, the time evolution of the particle position $x$ is described by the one-dimensional Langevin equation
\begin{equation}\label{eq:EoM}
\ddot{x}+\Gamma_0\dot{x}+\Omega_0^2x=\frac{1}{m}\left( \mathcal{F}_{\rm fluct}+F_{\rm ext}\right),
\end{equation}
where $\Omega_0\left/2\pi\right.\sim 125\rm kHz$ is the particle's angular frequency along the direction of interest, $\Gamma_0$ the friction coefficient and $F_{\rm ext}$ is an externally applied force. The random nature of the collisions does not only provide deterministic damping $\Gamma_0$ but also a stochastic force $\mathcal{F}_{\rm fluct}$, which thermalizes the energy of the nanoparticle. The fluctuation-dissipation theorem links the damping rate intimately to the strength of the stochastic force, $\mathcal{F}_{\rm fluct}(t)=\sqrt{2m\Gamma_0k_BT_0}\,\xi(t)$, with $T_0$, $k_B$  and $\xi(t)$ being the bath temperature, the Boltzmann constant, and white noise with $\langle\xi(t)\rangle = 0$ and $\langle\xi(t)\:\!\xi(t')\rangle=\delta(t-t')$.\\[-2ex]

\begin{figure}
\includegraphics[width=1.0\textwidth]{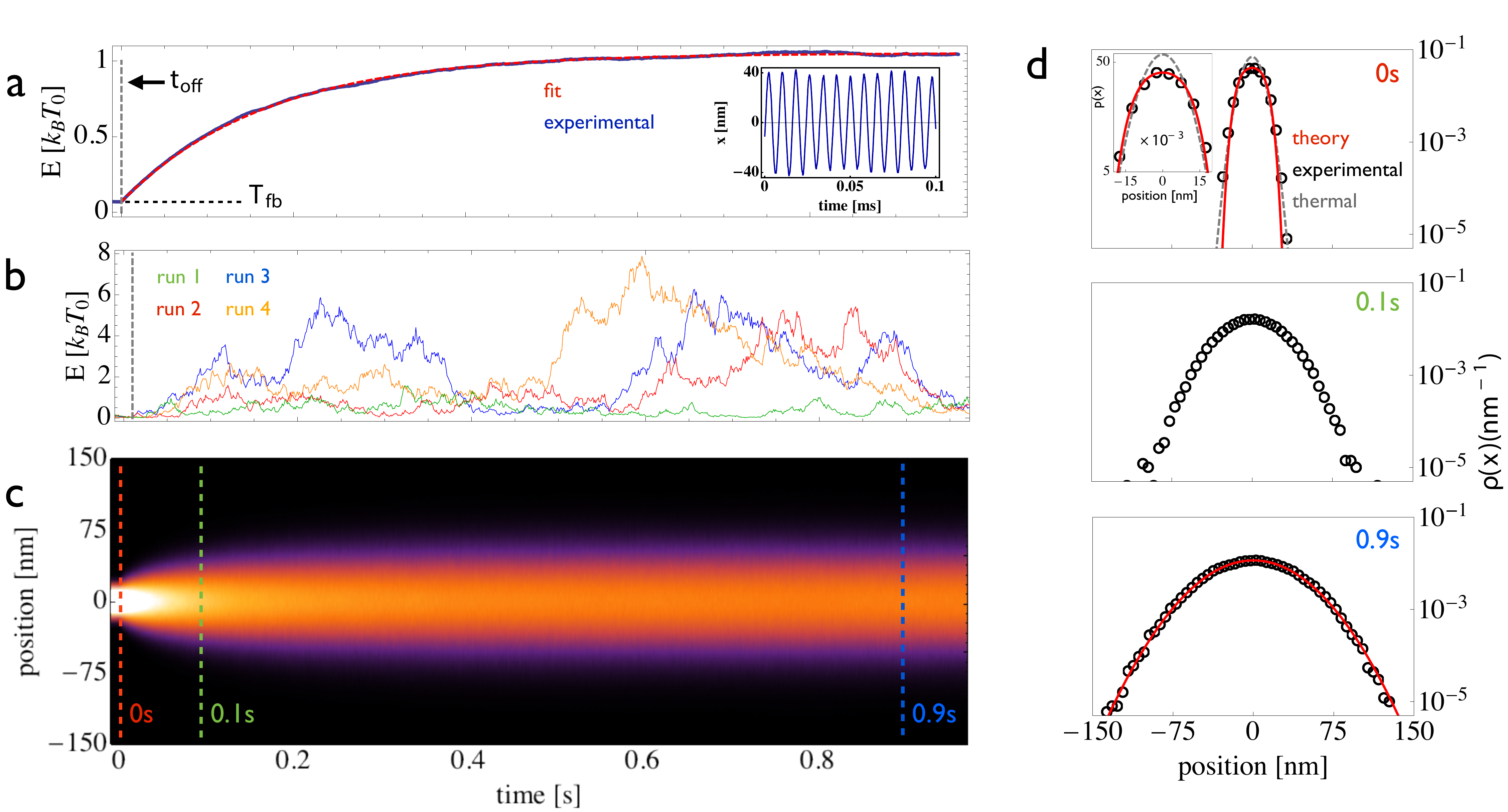}
\vspace{-1em}
\caption{
Relaxation from a non-equilibrium steady state generated by parametric feedback cooling.
The  initial effective temperature is $k_B T_{\rm fb}$. At time $t_{\rm off}$ the feedback is switched off and the particle energy relaxes to the equilibrium energy $k_BT_0$. (a) Time evolution of the average energy evaluated from $10^4$ individual experiments. The red dashed line is a fit according to Eq.~(\ref{eq:AverageEnergyRelaxation}). The inset shows that the particle oscillates with constant amplitude on short time scales.
(b) Four different realizations of the relaxation experiment. Each run yields a different trajectory and
the time it takes for the particle to acquire an energy of $k_BT_0$ deviates considerably from that of the average curve in (a).
(c) Time evolution of the position distribution shown as a density plot. (d) Position distributions evaluated at three different times. The distributions correspond to vertical cross-sections in figure (c).
The superimposed red curves are the theoretical distributions.
The initial distribution deviates significantly from a thermal equilibrium distribution with the same average energy (gray dashed line).
\label{fig:Figure1}}
\end{figure}

The total energy of the harmonically oscillating nanoparticle is given by
\begin{equation}\label{eq:HamiltonianHarmoniOscillator}
E(x,p)=\frac{1}{2}m\Omega_0^2 x^2+\frac{p^2}{2m}=\frac{1}{2}m\Omega_0^2 \bar{x}(t)^2,
\end{equation}
where  $x$ is the displacement form the trap center and  $p$ is the momentum. The second equality in the above equation follows from the slowly-varying amplitude approximation, $x(t)=\bar{x}\sin (\Omega_0 t)$, $\dot{\bar{x}}\ll \Omega_0 \bar{x}$. This approximation is well satisfied in our experiments since it takes many oscillation periods for the oscillation amplitude to change appreciably (see inset Fig.~\ref{fig:Figure1}a).\\[-2ex]

Applying a time-dependent external force $F_{\rm ext}$ for a sufficiently long time, the system is initially prepared in a non-equilibrium steady state with distribution $\rho_{\rm ss}(u, \alpha)$, which, in general, is not known analytically. Here, $u$ specifies the state of the system and $\alpha$ denotes one or several parameters that determine the initial steady state distribution, such as the strength of the external force. At time $t=t_{\rm off}$ the external force is switched off and we follow the evolution of the undisturbed system. In this relaxation phase (external force $F_{\rm ext}$ off) the dynamics satisfies detailed balance with respect to the equilibrium distribution $\rho_{\rm eq} \propto \exp(-\beta_0)E(u)$ at reciprocal temperature $\beta_0=1/k_{\rm B}T$. As shown be Evans and Searles \cite{Evans1994,Evans:2002gy} for thermostatted dynamics and by Seifert \cite{Seifert2012} for stochastic dynamics, the time reversibility of the underlying dynamics implies the transient fluctuation theorem  
\begin{equation}\label{eqn:FT_Y}
p(-\Y)\,/\,p(\Y)\;=\;e^{-\Y},
\end{equation}
holding for the {\em relative entropy change} 
\begin{equation}\label{eqn:Y}
\Y\,=\,\beta_0 Q+\Delta \phi\;.
\end{equation}
Here, $Q$ is the heat absorbed by the bath at reciprocal temperature $\beta_0$. Since no work is done on the system, the heat $Q$ exchanged along a trajectory of length $t$ starting at $u_0$ and ending at $u_t$ equals the energy lost by the system, $Q=-[E(u_t)-E(u_0)]$. The quantity $\Delta \phi=\phi(u_t)-\phi(u_0)$ is the difference of the trajectory-dependent entropy $\phi(u) = -\ln \rho_{\rm ss}(u, \alpha)$ \cite{Seifert:2005fu} between the initial and final state of the trajectory. Thus, $\Y$ is the change in relative entropy~\cite{Gaveau:1997ul}, or Kullback-Leibler divergence, between the initial steady state distribution and the equilibrium distribution observed along a particular trajectory. Note that the fluctuation theorem~(\ref{eqn:FT_Y}) holds for any time $t$ at which $\Y$ is evaluated and it is not required that the system reaches the equilibrium distribution at time $t$. The relative entropy change, which equals the dissipation function introduced by Evans and Searles for thermostatted dynamics \cite{Evans1994,Evans:2002gy,Carberry:2004fk}, is the logarithmic ratio of the probability to observe a particular trajectory and the probability of the corresponding time reversed trajectory \cite{Crooks:PRE:2000tc,Kawai:2007kc,Seifert2012}. As such, $\Y$ can be viewed as a measure of the irreversibility occurring during the relaxation process. 

From the detailed fluctuation theorem of Eq.~(\ref{eqn:FT_Y}) the integral fluctuation theorem
\begin{equation}\label{eqn:YYY}
\langle {\rm e}^{-\Y}\rangle\;=\; 1
\end{equation}
directly follows. Through Jensen's inequality, the convexity of the exponential function implies the second law-like inequality
\begin{equation}
\langle \Y \rangle\ge 0
\end{equation}
such that the average relative entropy change is non-negative. The average relative entropy change is related to the total entropy change of oscillator and bath together by \cite{Seifert2012}
\begin{equation}
\langle \Y \rangle=\Delta S_{\rm tot} + D(\rho_t\|\rho_{\rm ss})\; , 
\end{equation}
where $D(\rho_t\|\rho_{\rm ss}) $ is the relative entropy of the statistical state of the system at time $t$ with respect to the initial steady state distribution. Slightly modifying the definition of $\Y$ one can also derive a different but related integral fluctuation theorem~\cite{Jarzynski:JStatPhys:1999,Seifert:2005fu,Seifert2012}, from which the non-negativity of the total entropy change follows, $\Delta S_{\rm tot} \ge 0$, providing a direct link to the second law of thermodynamics. However, no detailed fluctuation theorem holds for this case. Analogous  fluctuation relations for the total entropy production have also been verified for two coupled systems kept in a non-equilibrium steady state by holding each system at a different temperature \cite{Ciliberto:2013gz,Koski:2013il}. For further discussion of the fluctuation theorem and the significance of $\Y$ see Supplementary Information.
\\[-2ex]

If the initial steady state distribution is an equilibrium distribution, $\rho_{\rm ss}(u, \alpha) = e^{-\beta [E(u)-F(\beta)]}$, corresponding to a temperature $T=1\left/k_B \beta\right.$ and with free energy $F(\beta)=-k_B T\,\ln\!\int \!du\,{\rm e}^{-\beta\,E(u)}$, the expressions become particularly simple and the fluctuation theorem for $\Y$ acquires a physically very transparent meaning. In this case, $\phi(u) =\beta[E(u)-F(\beta)]$, such that $\Y=(\beta_0-\beta) Q$ and the fluctuation theorem simplifies to $p(-Q)/p(Q)=\exp\{-(\beta_0 -\beta)Q\}$. Note that this particular fluctuation expression for the special case of transitions between equilibrium states has been obtained earlier~\cite{Jarzynski:2004ia} and was shown experimentally to hold also in the case of an ageing bath~\cite{GomezSolano:2011em}. As a consequence of this fluctuation relation for the heat,  the probability of observing energy flowing from the hotter bath to the colder system is exponentially small compared to the probability of observing energy transfer in the other direction. Since $Q$ is an extensive quantity, irreversibility for macroscopic systems is a direct consequence of the fluctuation theorem. The integral fluctuation theorem for the relative entropy change further implies that $(\beta_0-\beta) \langle Q\rangle \ge 0$, such that heat flows from hot to cold on the average, in line with the second law of thermodynamics.  

In the following, we experimentally investigate the fluctuation theorem \eqref{eqn:FT_Y} for two different initial non-equilibrium steady state distributions. The first steady state is generated by parametric feedback cooling ($\rm ss=fb$) and the second one by external modulation ($\rm ss=mod$) in addition to feedback cooling. In the case of parametric feedback cooling we enforce a non-equilibrium state  by applying a force $F_{\rm ext}=F_{\rm fb}$ to the oscillating particle through a parametric feedback scheme (c.f. Fig.~\ref{fig:Figure0})~\cite{Gieseler:2012bi}. The feedback $F_{\rm fb}=-\eta m\Omega_0 x^2\dot{x}$ adds a cold damping $\Gamma_{\rm fb}$ to the natural damping $\Gamma_0$. This is different from thermal damping, where an increased damping is accompanied by an increase in fluctuations. Since parametric feedback adds an amplitude dependent damping $\Gamma_{\rm fb} \propto x^2$, oscillations with a large amplitude experience a stronger damping than oscillations with a small amplitude. As a consequence, the position distribution is non-Gaussian and assumes the form (see Supplementary Information)
\begin{equation}\label{eqn:SteadyStateDistFB}
\rho_{\rm fb}(x,\alpha)= \sqrt{\frac{\beta_0 m\Omega_0^2(4+\alpha m\Omega_0^2x^2)}{8\pi^3}}\,
\frac{\exp\left[-\frac{\beta_0(4+\alpha m\Omega_0^2x^2)^2}{32\alpha}\right]}{\text{erfc}\left(\sqrt{\beta_0/\alpha}\right)}
\text{K}_{1/4}\left[\frac{\beta_0(4+\alpha m\Omega_0^2x^2)^2}{32\alpha}\right],
\end{equation}
where  $\alpha=\eta\left/m \Gamma_0\Omega_0\right.$, and $\text{erfc}$ and $\text{K}_{1/4}$ are the complementary error function and a generalized Bessel function of the second kind, respectively. In analogy to the thermal equilibrium temperature of the harmonic oscillator, we define an effective temperature $T_{\rm fb}=\langle E \rangle_{\rm fb}/k_B$ of the system. Here $\langle E \rangle_{\rm fb}$ denotes the average energy with feedback on. Using the  distribution ~(\ref{eqn:SteadyStateDistFB}) to calculate the average energy we find the effective temperature 
%
\begin{equation} 
T_{\rm fb} =T_0\left\{2\sqrt{\frac{\beta_0}{\alpha}}\frac{e^{-\beta_0\left/\alpha\right.}}{\sqrt{\pi}\,\text{erfc}\left(\sqrt{\beta_0\left/\alpha\right.}\right)}-2\frac{\beta_0}{\alpha}\right\}\approx \; \sqrt{\frac{4m\Gamma_0\Omega_0T_0}{\pi k_B\eta}},
\end{equation}
where the approximation holds for $T_{\rm fb}/T_0\ll 1$.

At time $t=t_{\rm off}$ the feedback is switched off and the system relaxes back to the thermal equilibrium distribution at temperature $T_0$. The experimental data for this relaxation process are shown in Fig.~\ref{fig:Figure1}c and d. Without the feedback, the collisions with the surrounding molecules are no longer compensated and the oscillator energy increases. Exploiting that at low friction the oscillator energy changes slowly, one finds from Eqs. \eqref{eq:EoM} and \eqref{eq:HamiltonianHarmoniOscillator} that the time evolution of the energy is governed by $\dot E = -\Gamma_0 (E  -k_{\rm B}T_0)+\sqrt{2E \Gamma_0 k_{\rm B}T_0}\xi(t)$. An average over noise then yields the differential equation $\langle \dot E \rangle=-\Gamma_0 (\langle E \rangle  - k_{\rm B}T_0)$, which implies that the average energy of the oscillator relaxes exponentially to the equilibrium value $k_B T_0$, 
\begin{equation}\label{eq:AverageEnergyRelaxation}
\langle E(t)\rangle=k_BT_0+k_B (T_{\rm ss}-T_0)e^{-\Gamma_0 t}\;,
\end{equation}
where $T_{\rm ss}$ denotes an arbitrary initial steady-state temperature, for example $T_{\rm fb}$.\\[-2ex]

To verify this equation, we repeat the relaxation experiment $10^4$ times. Each time the same initial distribution $\rho_{\rm fb}(u_0, \alpha)$ is established by parametric feedback and, after switching off the feedback, the system is followed  as it evolves from $u_0$ to $u_t$ within time $t$. Along each $\sim 1s$ trajectory we sample the particle position at a rate of $625\rm kHz$ and from integration over 64 successive position measurements we obtain the energy at a rate of $9.8\rm kHz$. In Fig. \ref{fig:Figure1}a we show the average over the individual time-traces together with a fit to Eq. \eqref{eq:AverageEnergyRelaxation}. Equilibrium is reached after a time of the order of $\tau_0=1/\Gamma_0=0.17\rm \; s$. According to Eq.~\eqref{eq:AverageEnergyRelaxation} and the data shown in Fig. \ref{fig:Figure1}, the average energy of the particle increases monotonically. However, due to the small size of the particle, the fluctuating part $\sqrt{2E \Gamma_0 k_{\rm B}T_0}\xi(t)$ is comparable to the deterministic part $-\Gamma_0 (E  -k_{\rm B}T_0)$ and hence an individual trajectory can be quite different from the ensemble average of Eq. \eqref{eq:AverageEnergyRelaxation}. Figure \ref{fig:Figure1}b shows four realizations of the relaxation experiment. Each particle trajectory $x(t)$ results from switching off the feedback at  initial time $t=t_{\rm off}$.\\[-2ex]

\begin{figure}
\includegraphics[width=0.8\textwidth]{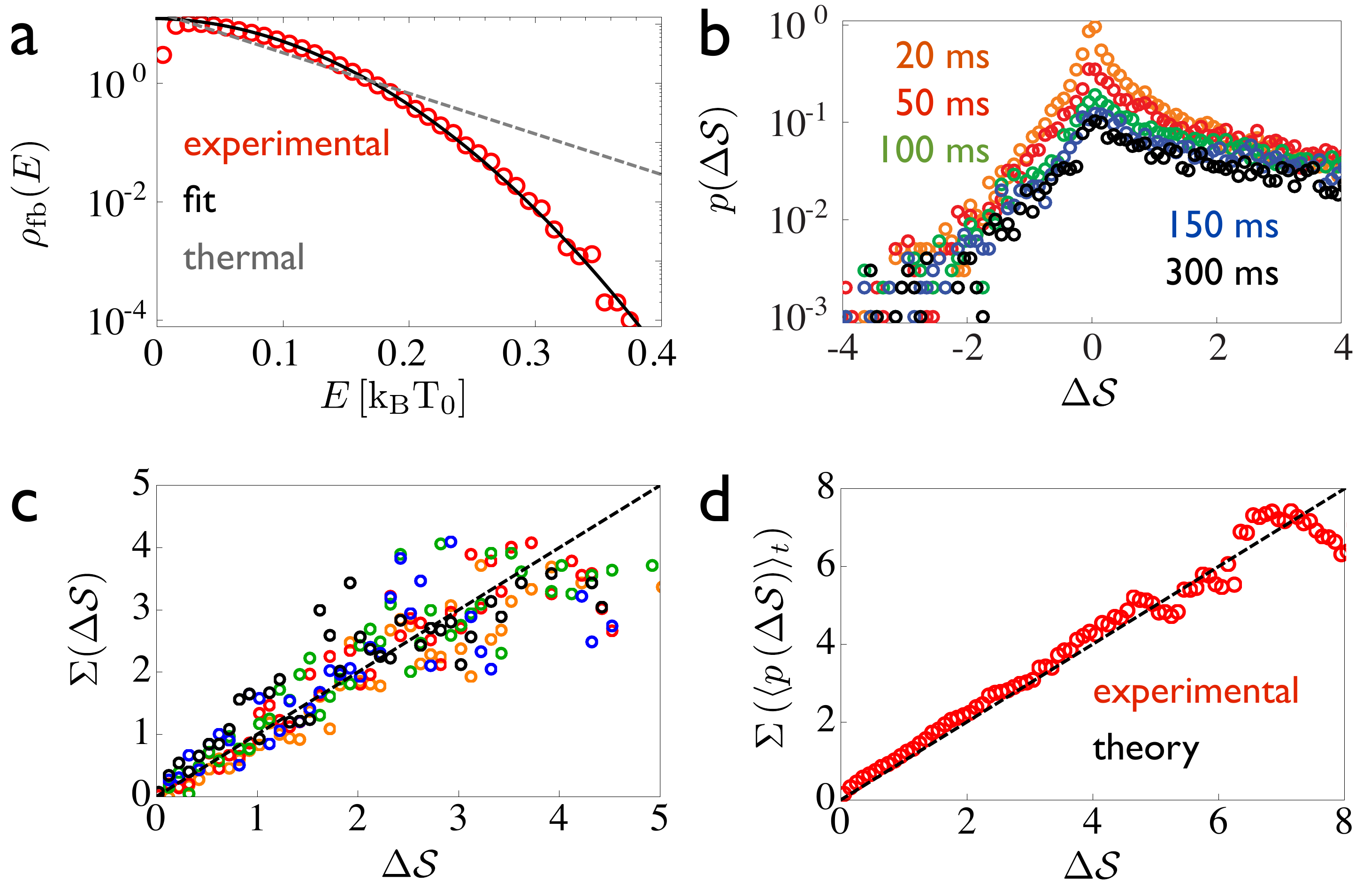}
\vspace{-1.5em}
\caption{
Fluctuation theorem for the relaxation experiment  in Fig.~\ref{fig:Figure1}.
(a) Energy distribution with feedback on (red circles).  The black solid curve is a fit according to Eq.~(\ref{eqn:SteadyStateDistFBEnergy}).
Large amplitude oscillations experience stronger damping and are therefore suppressed relative to an equilibrium distribution (gray dashed line).
(b) Probability density $p(\Y)$ evaluated for different times after switching off the feedback.
(c) The function $\Sigma(\Y)$ evaluated for the distributions shown in (b). 
(d) The function $\Sigma$ evaluated for the time averaged distributions $\langle p(\Y)\rangle_t$. The data are in excellent agreement with the fluctuation theorem of Eq.~(\ref{eqn:FT_Y}) (black dashed line).
\label{fig:Figure2}}
\end{figure}

The $10^4$ trajectories allow us to evaluate the distributions $p_{\rm fb}(\Y)=\langle \delta[\Y-\Y(u_t)] \rangle_{\rm fb}$ for different times $t$. Here, the subscript '$\rm fb$' denotes the average over the initial distributions obtained under the action of feedback. For this initial non-equilibrium steady state the energy distribution is calculated analytically as (see Supplementary Information)
\begin{equation}\label{eqn:SteadyStateDistFBEnergy}
\rho_{\rm fb}(E,\alpha)\;=\; \sqrt{\frac{\alpha\beta_0}{\pi}}
\,\frac{\exp\left(-\beta_0/\alpha\right)}{{\rm erfc}\left(\!\sqrt{\beta_0/\alpha}\right)}\:\exp\!\left(\!-\beta_0 \left[E+\frac{\alpha}{4} E^2\right]\right).
\end{equation}
This distribution has the form of a Boltzmann-Gibbs distribution for the generalised energy $E+\alpha E^2/4$, where the term $\alpha E^2/4$ arises from the feedback and strongly penalises high energy states.
It is consistent with the phonon number distribution of an optomechanical system with a quadratic coupling term \cite{Nunnenkamp:2010gj}.
Inserting the above distribution into Eq.~(\ref{eqn:Y}) we find that for the relaxation from $\rho_{\rm fb}$ the relative entropy change is given by $\Y=\beta_0\alpha\left(E_t^2-E_0^2\right)/4$. In this case, the integral fluctuation theorem implies that $\langle \Delta E^2 \rangle \ge 0$, i.e., the average of the squared energy does not decrease during the relaxation process.
Figure~\ref{fig:Figure2}a shows the measured steady state distribution of the energy in excellent agreement with the prediction of Eq.~\eqref{eqn:SteadyStateDistFBEnergy}.
For small energies, the measured distribution features a small dip caused by measurement noise.
For comparison, we also show the corresponding equilibrium distribution with the same average energy (gray dashed line).
It is evident that it deviates strikingly from the true distribution $\rho_{\rm fb}(E,\alpha)$. In Fig.~\ref{fig:Figure2}b we plot the distributions $p_{\rm fb}(\Y)$ for different times $t$.
They become increasingly asymmetric for long times, with higher probabilities for positive $\Y$ and lower probabilities for negative $\Y$. To test the fluctuation theorem \eqref{eqn:FT_Y} for our measurements we define
\begin{equation}
\label{eq:FluctuationFunction}
\Sigma(\Y)=\ln \left[\frac{p(\Y)}{p(-\Y)}\right]= \Y,
\end{equation}
where $\Sigma(\Y)$ is predicted to be time-independent. Using the distributions for $\Y$ shown in Figure \ref{fig:Figure2}b we compute $\Sigma(\Y)$ and show the resulting data in Fig.~\ref{fig:Figure2}c. Since the fluctuation theorem~\eqref{eqn:FT_Y} is time-independent, we evaluate the time-average for each $\Y$ in Fig.~\ref{fig:Figure2}c and render it in the plot shown in Fig.~\ref{fig:Figure2}d. The averaging improves the statistics and leads to excellent agreement with the fluctuation theorem for $\Y$.
The offset for small $\Y$ results from measurement noise.
\\[-2ex]

\begin{figure}
\includegraphics[width=1.0\textwidth]{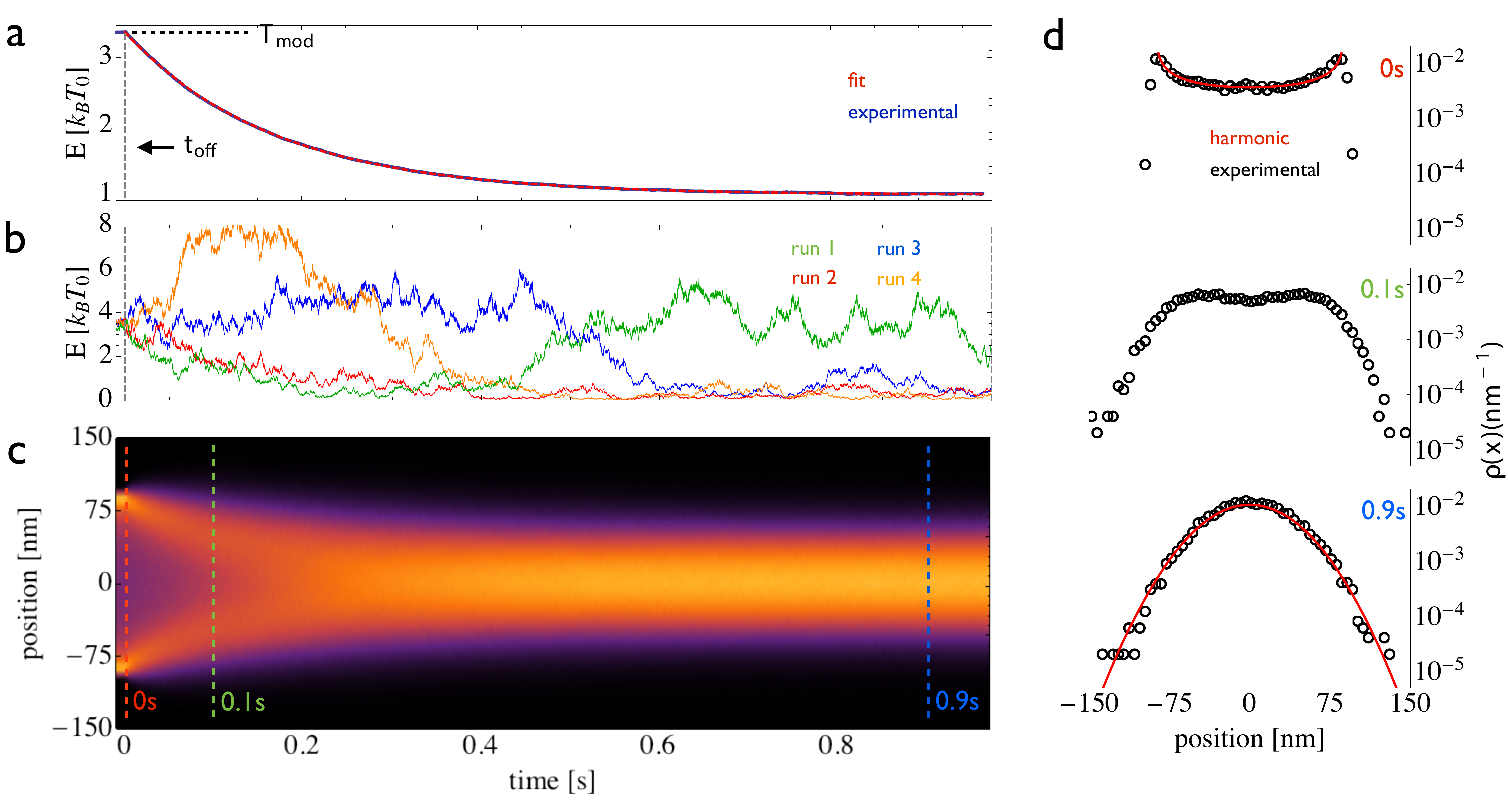}
\vspace{-1em}
\caption{
Relaxation from a non-equilibrium steady state generated by external parametric modulation. The  initial effective temperature is $k_B T_{\rm eff}$. At time $t_{\rm off}$ the feedback is switched off and the particle energy relaxes to the equilibrium energy $k_BT_0$. 
(a) Time evolution of the average energy evaluated from repeated individual experiments. The red dashed line is a fit according Eq.~(\ref{eq:AverageEnergyRelaxation}). 
(b) Four different realizations of the relaxation experiment. Each run yields a different trajectory and
the time it takes for the particle to acquire an energy of $k_BT_0$ deviates considerably from that of the average curve (a).
(c) Time evolution of the position distribution shown as a density plot. 
(d) Position distributions evaluated at three different times. The distributions correspond to vertical cross-sections in figure (c).
The superimposed red curves are the theoretical distributions.
The initial distribution features a sharply peaked double-lobe distribution, characteristic for a harmonic oscillator at constant energy.
As the system evolves, the two peaks smear out and merge into a single Gaussian distribution.
\label{fig:Figure3}}
\end{figure}

The experimental scheme introduced here allows us to  study non-equilibrium processes for arbitrary initial states and for arbitrary transitions between states. To demonstrate that the fluctuation theorem holds for arbitrary non-equilibrium initial states, we apply an external harmonic drive signal in addition to the parametric feedback as illustrated in Fig.~\ref{fig:Figure0}. The harmonic drive generates a force $F_{\rm mod}=\epsilon m\Omega_0^2\cos(\Omega_{\rm mod}t)x$ acting on the nanoparticle, with modulation frequency $\Omega_{\rm mod}/2\pi=249\rm kHz$ and modulation depth $\epsilon=0.03$. Modulation at $\Omega_{\rm mod}$ brings the particle into oscillation at frequency $124.5\rm kHz$ and amplitude $\bar x$. The resulting steady state position distribution $\rho_{\rm mod}(x)$ deviates strongly from an equilibrium Gaussian distribution and resembles  the characteristic double-lobe function 
\begin{equation}
\rho_{\rm mod}(x)=\frac{\pi^{-1}}{\sqrt{{\bar x}^2-x^2}}
\end{equation}
of a harmonic oscillator with constant energy. As in the previous experiment, at $t=t_{\rm off}$ the modulation and the feedback are switched off, and  the nanoparticle dynamics is measured during relaxation.
Figure~\ref{fig:Figure3} shows the relaxation of the particle's average energy and the evolution of the position distribution.\\[-2ex]

Due to the additional driving, the average initial energy is larger than the thermal energy $k_B T_0$. After the driving is switched off, the average energy relaxes exponentially to the equilibrium value according to Eq.~\eqref{eq:AverageEnergyRelaxation}.
As in the previous experiment, individual realizations of the switching experiment differ significantly from the average (Fig.~\ref{fig:Figure3}b). As the system relaxes, the two lobes of the initial position distribution broaden until they merge into a single Gaussian peak corresponding to temperature $T_0$.\\[-2ex]

\begin{figure}
\includegraphics[width=0.8\textwidth]{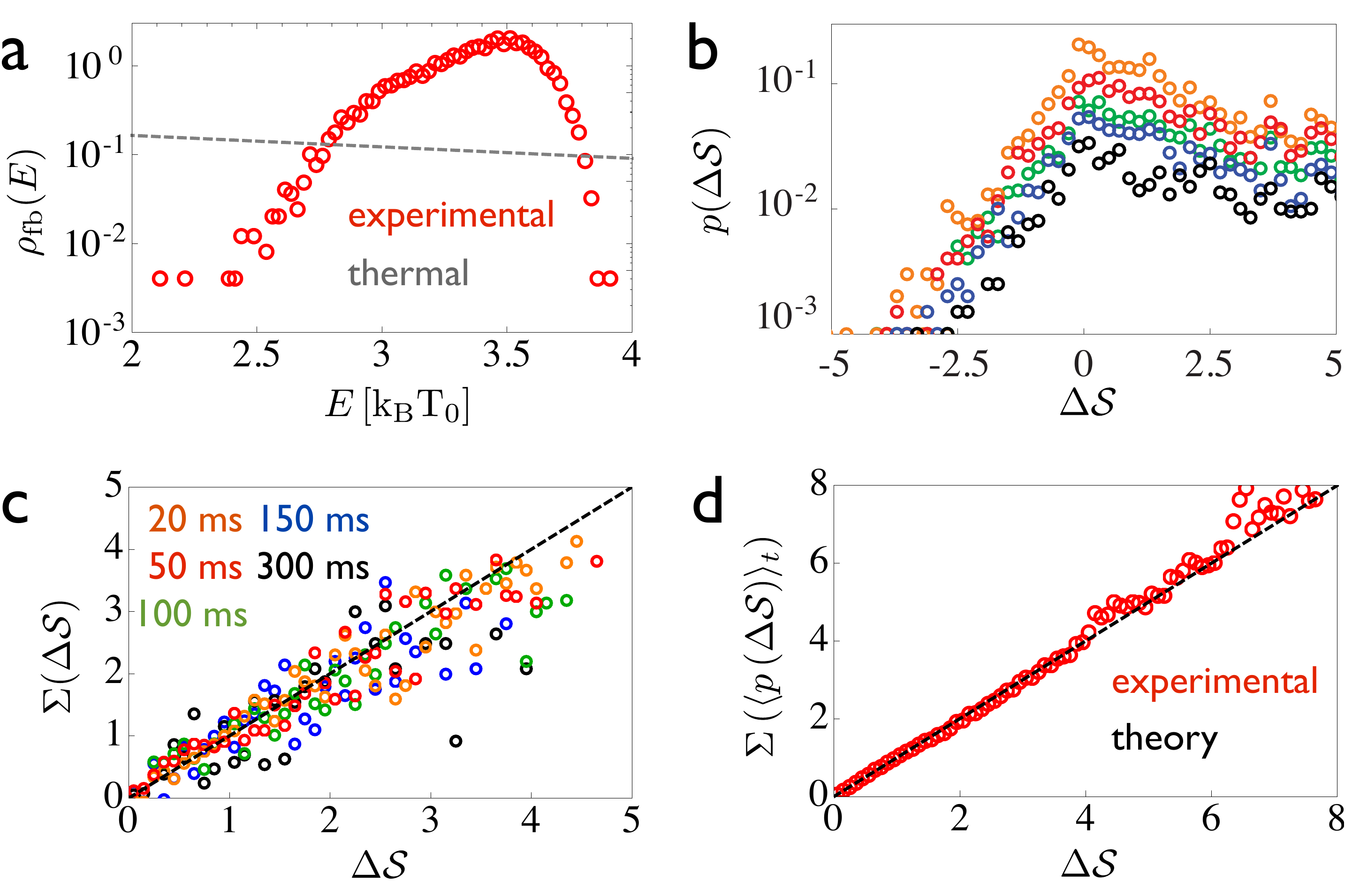}
\vspace{-1.5em}
\caption{
Fluctuation theorem for the relaxation experiment  of Fig.~\ref{fig:Figure3}.
(a) The energy distribution with external modulation on (red circles) differs significantly from an equilibrium distribution with identical average energy (gray dashed line).
(b) Probability density $p(\Y)$ evaluated for different times after switching off the modulation.
(c) The function $\Sigma(\Y)$ evaluated for the distributions shown in (b). 
(d) The function $\Sigma$ evaluated for the time averaged distributions $\langle p(\Y)\rangle_t$. The data are in excellent agreement with the fluctuation theorem of Eq.~(\ref{eqn:FT_Y}) (black dashed line).\label{fig:Figure4}}
\end{figure}

In the case of parametric modulation, the form of the initial energy distribution $\rho_{\rm mod}(E)$ is not known analytically  and therefore needs to be determined experimentally. Using the measured initial distribution together with the energies $E_0$ and $E_t$ evaluated at times $0$ and $t$, respectively, we calculate $\Y=\beta_0 Q+\Delta \phi$. Figure~\ref{fig:Figure4}a shows the initial energy distribution $\rho_{\rm mod}(E)$, which has a narrow spread around a non-zero value and therefore differs significantly from a thermal distribution with identical effective temperature (gray dashed line).
The measured distributions of $\Y$ evaluated at different times after switching off the modulation are shown in Fig.~\ref{fig:Figure4}b. As before, we use the distributions $p(\Y)$ to evaluate $\Sigma(\Y)$ and plot it in Fig.~\ref{fig:Figure4}c. To reduce the variance we  time-average the distributions $p(\Y)$ and plot the corresponding $\Sigma$ function in Fig.~\ref{fig:Figure4}d. As in the previous experiment, we find excellent agreement with the theory (black dashed line), providing solid experimental validation of the fluctuation theorem~(\ref{eqn:FT_Y}) valid for initial steady states that are out of equilibrium. \\

In conclusion, we have experimentally demonstrated the validity of a fluctuation theorem for the relaxation from a non-equilbrium state towards equilibrium. The theorem holds for the relative entropy change $\Y$, which is related (but not identical) to the total entropy production. Using a levitated nanoparticle in high vacuum we have verified the fluctuation theorem for different initial non-equilibrium states, demonstrating that this theoretical framework can be used to understand fluctuations in nanoscale systems. Our experimental approach allows us to measure the dynamics of a nanoparticle during relaxation from an arbitrary initial state and to study its statistical properties. We succeeded in deriving an analytic expression for the non-equilibrium steady state under the action of a feedback force and demonstrated excellent agreement with experimental data. The here presented experimental framework naturally extends to the study of transitions between arbitrary steady states and to quantum fluctuation theorems, similar to recent proposals for trapped ions  \cite{Huber:2008go,Campisi:2011ka}. We envision that our approach of using highly controllable nanomechanical oscillators will open up experimental and theoretical studies of fluctuation theorems in complex settings, which arise, for instance, from the interplay of thermal fluctuations and nonlinearities \cite{Gieseler:2013uu} where detailed balance does not hold \cite{Dykman:1979vj,Dykman:2011ug}. Furthermore, it serves as an experimental simulator platform in analogy to quantum simulators based on ultracold gases, superconducting circuits or trapped ions~\cite{cirac12}. \\

\newpage



\end{document}